\documentclass[aps,prl,twocolumn,showpacs]{revtex4-1}
\usepackage{braket}
\usepackage{upgreek}
\usepackage{graphicx}
\usepackage{amsmath}
\begin{document}

\title{Testing multi-photon interference on a silicon chip
}

\author{Bryn A. Bell}
\email{bryn.bell@physics.ox.ac.uk}
\author{Guillaume S. Thekkadath}
\author{Ian A. Walmsley}

\affiliation{Department of Physics, University of Oxford, Clarendon Laboratory, Parks Road, Oxford, OX1 3PU, United Kingdom}

\author{Renyou Ge}
\author{Xinlun Cai}
\email{caixlun5@mail.sysu.edu.cn}
\affiliation{State Key Laboratory of Optoelectronic Materials and Technologies and School of Electronics and Information Technology, Sun Yat-sen University, Guangzhou 510000, China}





\begin{abstract}
Multi-photon interference in large multi-port interferometers is key to linear optical quantum computing and in particular to boson sampling. Silicon photonics enables complex interferometric circuits with many components in a small footprint and has the potential to extend these experiments to larger numbers of interfering modes. However, loss has generally limited the implementation of multi-photon experiments in this platform. Here, we make use of high-efficiency grating couplers to combine bright and pure quantum light sources based on ppKTP waveguides with silicon circuits. We interfere up to 5 photons in up to 15 modes, verifying genuine multi-photon interference by comparing the results against various models including partial distinguishability between photons.
\end{abstract}
\maketitle


Multi-photon interference in large multi-port interferometers is central to optical quantum computing~\cite{obrien07}. In particular it has been shown that boson sampling, where single photons propagate through a Haar-random, static interferometer followed by single photon detectors, is hard to emulate with classical computation~\cite{aaronson11}. More recently, scattershot boson sampling (SBS) and Gaussian boson sampling (GBS) have also been shown to be computationally hard~\cite{Lund14, Hamilton17}. These protocols use squeezed light or probabilistically-generated photon-pairs instead of single photons, which can be generated using spontaneous parametric downconversion (SPDC) or four-wave mixing (FWM). This is an experimentally appealing route towards the demonstration of a quantum computational advantage because these sources are readily available and could feasibly be scaled to many identical sources operated in parallel, particularly thanks to progress in waveguided and integrated sources~\cite{Spring17, Harder16, Wang18}.

There is still a gulf between experimental efforts, which have demonstrated interference between up to 5 photons in up to 16 optical modes~\cite{Wang17,Zhong18,Bentivegna15, HWang18}, and the numbers required for a convincing demonstration of a quantum advantage over classical super-computers, estimated at 50 photons in over a thousand modes~\cite{Clifford18}. As well as the need to increase the number of photon sources, it is challenging to implement such a large interferometer while keeping photon loss sufficiently low. To date, the most complex linear optical circuits have been realised in silicon photonics, which allows a high component density and scalable fabrication technology~\cite{Wang18, Sun13, Shen17, Harris17, Qiang18}. However, silicon suffers from relatively high losses, particularly the coupling loss to optical fiber due to mode mis-match; this has largely restricted quantum photonics in silicon to two photon interference experiments~\cite{Qiang18, Silverstone14, Silverstone15}, with a few examples of higher photon-numbers~\cite{Harada11, Zhang16, Faruque18, Bell18, Paesani19}. These examples have made use of on-chip FWM to generate photon-pairs, which has the advantage of close integration between the sources and the interferometer, avoiding loss from coupling the photons onto the chip. However, these sources tend to be at a disadvantage compared to SPDC sources in terms of count rates and interference quality, and suffer from internal losses and from parasitic nonlinearities such as two photon absorption, self-phase modulation, and free-carrier effects~\cite{Husko13}. In particular for GBS and related applications requiring relatively high squeezing levels, parasitic nonlinearities are likely to prove a barrier.

Otherwise, multiphoton experiments have made use of bulk-optic SPDC or quantum dot sources. For SPDC sources in particular, strategies have been developed to simultaneously optimise the collection efficiency and the purity of the photons, generating highly indistinguishable photons while avoiding the need for narrow filtering~\cite{Grice01, MeyerScott17}. Interferometers are realised in bulk-optics or silica waveguides, which are well mode-matched to fiber. This approach makes use of individually well-optimised and low-loss components, but due to their bulkiness it is unlikely these approaches can be extended to very large numbers of interfering modes~\cite{Wang17, Zhong18, Bentivegna15}.


\begin{figure}[t]
	\centering
	\includegraphics[width=\columnwidth]{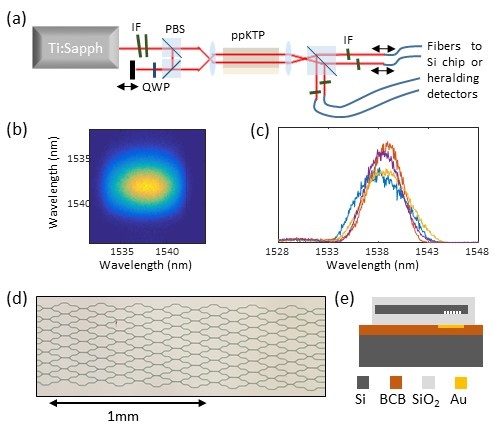}
        \vspace{-10pt}
	\caption{(a) SPDC source setup, comprising Ti:Sapphire laser, interference filters (IF), polarizing beam-splitters (PBS), quarter waveplate (QWP), ppKTP waveguides, and fiber collection stages. (b) Measured joint spectral intensity of source A. (c) Normalized spectra of the generated photons, red and blue lines: signal and idler for source A; purple and yellow lines: signal and idler for source B. (d) Microscope image of a 15x15 interferometer realised as a network of directional couplers on a silicon chip. (e) Vertical structure of the chip, showing Si waveguide and grating coupler with a SiO$_2$ cladding, Au reflector below the grating coupler, bonded to a Si carrier by BCB.}\label{fig1}
 \vspace{-5pt}
\end{figure}


In this work we demonstrate multi-photon interference between up to 5 photons in 13 and 15 mode Haar-random interferometers on a silicon chip, each occupying less than 1mm$^2$. The photons are generated using SPDC in two waveguides on the same periodically-poled potassium titanyl phosphate (ppKTP) crystal. ppKTP waveguides can generate pure and indistinguishable photons with high collection efficiency, and very high levels of squeezing are available with moderate pump power~\cite{Harder16}. High-efficiency grating couplers, with an apodized grating structure and metal back-reflectors, are used to couple the photons on and off the silicon chip with low-loss. On the silicon chip, the modes are combined using the interferometer design proposed in~\cite{Clements16}, which can realise any unitary transformation while minimising the propagation distance through the interferometer. We compare our results to models which include partial distinguishability - the effects of partial distinguishability on multiphoton interference are non-trivial, but in general mean that an $N$ photon experiment can be approximated by a mixture of $k$ photon interference effects, where $k<N$, thereby reducing the hardness of classical computation~\cite{Tichy15, Renema18}. Experimental implementations are often verified by showing that the results are closer to the perfectly interfering case of indistingushable photons than to the case where all photons can be distinguished from each other~\cite{Spagnolo14}, neglecting the possibility that the experiment is well-described by partially interfering photons where genuine $N$ photon interference is either absent or negligible compared to lower orders. Here we find that our results are best described by models that include all orders of interference, and hence conclude that genuine $N$ photon interference occurs. This work demonstrates that by properly engineering the coupling to fibre it is possible to combine complex interferometric circuits in silicon with bright and pure SPDC sources, and suggests that in addition to the route of full integration, with on-chip sources and detectors, a hybrid approach is viable which interfaces the best sources, interferometers, and detectors for a given application.

The source setup is depicted in Fig.~\ref{fig1}(a). A Ti:Sapphire laser generates pump pulses at 769nm with 76MHz repetition rate and is filtered to a bandwidth of 2nm by a pair of angle-tuned interference filters. The pulses are split into two paths by a polarizing beamsplitter then coupled to two near-identical waveguides in the same ppKTP crystal, where type II SPDC occurs, generating degenerate photon pairs at 1538nm. A movable delay in one pump path is used to adjust the arrival time between the two sources. The generated photons are separated by polarization, filtered, and collected into fiber. Translation stages on two of the fiber ports adjust the timing between pairs of photons from the same source.

An advantage of SPDC in ppKTP in this configuration is that the spectral correlation between the signal and idler from each source is minimised, so that the photons arrive at the interferometer in a near pure state without the need for narrow filtering~\cite{Grice01}. The filters after the sources have a 10nm bandwidth, which is wide enough to transmit the central peak of the photons' spectra while blocking the pump light, the side-lobes resulting from the phase-matching sinc function, and a broad background of photons resulting from SPDC involving higher-order spatial modes. Fig.~\ref{fig1}(b) shows the joint spectral intensity of source A after filtering, measured using a time-of-flight spectrometer based on dispersion compensating fiber followed by single photon detectors. No spectral correlation is apparent. Fig.~\ref{fig1}(c) shows the 4 individual spectra for signal and idler photons from the 2 sources - here, slight differences in central wavelength and bandwidth are apparent, resulting in some distinguishability. The difference in bandwidth between signal and idler originates from the fact that their group velocities in ppKTP are not exactly symmetrically spaced to either side of that of the pump pulse. 

\begin{figure}[t]
	\centering
	\includegraphics[width=0.85\columnwidth]{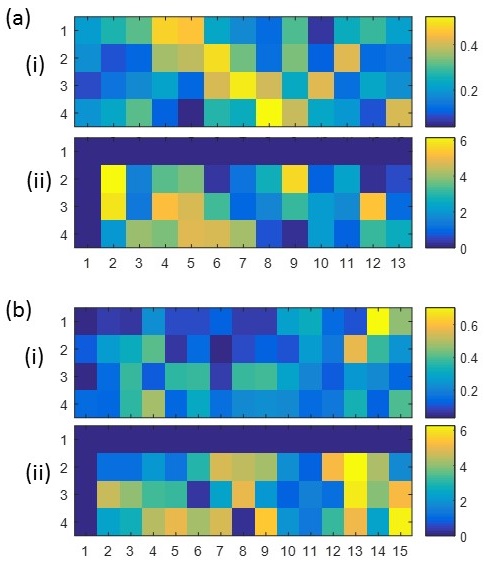}
        \vspace{-10pt}
	\caption{Transfer matrices for (a) the 13 and (b) the 15 mode interferometer. (i) Absolute value and (ii) phase in radians.}\label{fig2}
 \vspace{-5pt}
\end{figure}

\begin{figure*}[t]
	\centering
	\includegraphics[width=2\columnwidth]{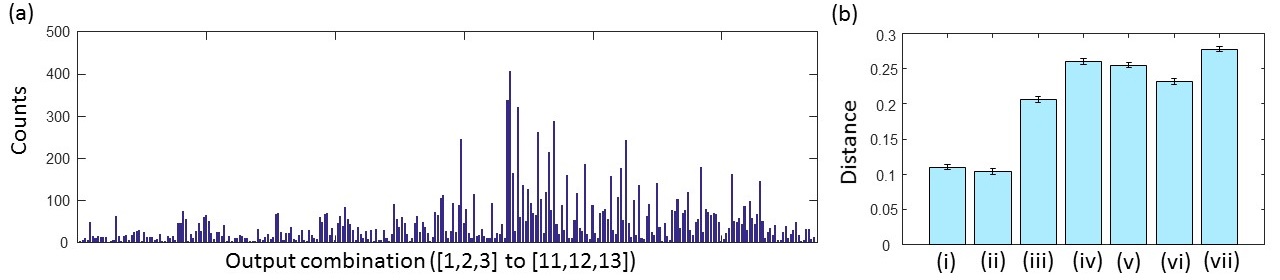}
        \vspace{-10pt}
	\caption{(a) Distribution of counts for 3 photons in a 13x13 interferometer. (b) Distance from various theoretical models: (i) ideal indistinguishable photons; (ii) partially distinguishable photons; (iii) truncated at two photon interference; (iv)-(vi) respectively the first, second, and third photon is completely distinguishable while the other two interfere perfectly; (vii) all distinguishable photons. Error bars are based on Monte Carlo simulation of Poissonian count statistics.}\label{fig3}
 \vspace{-5pt}
\end{figure*}

Fig.~\ref{fig1}(d) shows a microscope image of the 15x15 mode interferometer and Fig.~\ref{fig1}(e) shows the vertical structure of the chip. The interferometer was fabricated by e-beam lithography and inductively coupled plasma etching on 220nm silicon-on-insulator. It comprises a network of 105 directional couplers (DC) in a square mesh. This geometry can realise any unitary transformation given particular choices of the DC transmission coefficients and of the phase-shifts between DCs~\cite{Clements16}. It is also possible to directly create a Haar-random unitary by choosing the transmission coefficients according to particular probability distributions, which are dependent on the position of the DC within the network~\cite{Burgwal17, Russell17}. Meanwhile all phase-shifts are chosen from uniform random distributions. Here, the DC coefficients were chosen using these distributions then translated into the corresponding lengths of the coupling region, while the phase-shifts were randomised by slight differences in path-length between DCs. The photons are coupled between the silicon chip and a fiber array using high-efficiency grating couplers. Here, the grating lines are made up of a pattern of nanoscale holes, which are varied in size to appodize the grating and match the emission to the fiber mode~\cite{Ding14, Luo18}. Metal bottom mirrors are also used to prevent emission from below the gratings into the substrate. To fabricate the bottom mirrors, Ti and Au were first patterned above the grating couplers, then the chip was flip-bonded to a Si wafer using benzocyclobuten (BCB). The original substrate of the chip was removed to allow access from above in the new orientation~\cite{supp}. At the optimum wavelength of 1547nm the grating couplers achieve a coupling loss of approximately 1dB; however, at the photons' wavelength of 1538nm this is increased to 1.5dB. Values as low as 0.58dB have been achieved elsewhere, approaching the coupling efficiencies seen between fibre and silica waveguides~\cite{Ding14}. The output fibers are connected to superconducting nanowire single photon detectors having a detection efficiency of approximately 85\%.

For the 13 mode (15 mode) interferometer, a 4x13 (4x15) transfer matrix was characterized, since in the following experiments only up to 4 input modes were used. First the absolute values of the matrix were measured by injecting single photons into each input sequentially and monitoring the counts at every output. Then two photon interference was used to retrieve the complex phases of the matrix~\cite{Laing12, Spring13}. Pairs of photons from source A were injected into two input modes, and their relative delay was scanned to locate a Hong-Ou-Mandel (HOM) dip. The two photon coincidence rates were monitored for all combinations of two output modes, all of which contain a HOM dip or anti-dip with a visibility depending on the internal phases of the interferometer. This was repeated for the 6 possible choices of two out of four inputs. This provided a set of 468 (630) HOM visibilities; some contain a larger statistical uncertainty than others because of the variation in count rates between choices of input/output combination. A maximum likelihood optimization was then used to retrieve the phases based on all of the visibilities and their corresponding uncertainties. The results are shown in Fig.~\ref{fig2}. Since multi-photon interference is not sensitive to external phase-shifts on the inputs and outputs of the interferometer, the phases of the first row and column are set to zero. A continuous-wave laser at 1539nm was also used to make a classical measurement of the overall transmission through the chips, averaged over the first 4 inputs. The transmission was 39$\%$ for the 13 mode interferometer and 35$\%$ for the 15 mode interferometer.

Three photons were then injected into the 13 mode interferometer, with source A generating a photon pair in the first two inputs and source B used to herald a single photon which was injected into the third input. The distribution of counts over the 286 possible collision-free output combinations is shown in Fig.~\ref{fig3}(a), after 2 hours of integration time at a count rate of 1.8Hz. The ideal distribution was also calculated, by taking permanents of 3x3 sub-matrices from the transfer matrix. The fidelity between experimental and ideal distributions is $F=\left(\sum_j \sqrt{p_j.q_j}\right)^2=0.98$, with $p_j$ and $q_j$ the ideal and measured distributions normalised to sum to 1.

\begin{figure*}[t]
	\centering
	\includegraphics[width=2.1\columnwidth]{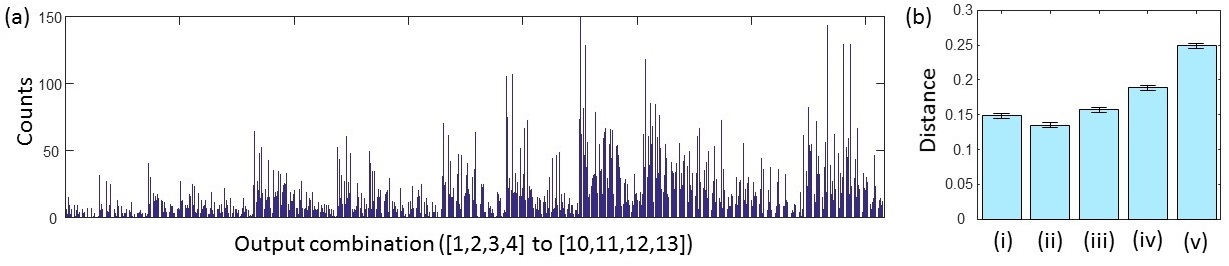}
        \vspace{-10pt}
	\caption{(a) Distribution of counts for 4 photons in a 13x13 interferometer. (b) Distance from various theoretical models: (i) ideal indistinguishable photons; (ii) partially distinguishable photons; (iii) truncated at three photon interference; (iv) truncated at two photon interference; (v) all distinguishable photons.}\label{fig4}
 \vspace{-5pt}
\end{figure*}

A more sensitive measure is the distance,
\begin{equation}
D=\frac{1}{2}\sum_j |p_j-q_j|=0.097,
\end{equation}
which varies from 0 when the distributions are identical to 1 when they do not overlap. We use $D$ as a measure to compare the experimental results to several theoetical models containing partial distinguishability, in particular aiming to rule out models which do not contain genuine 3 photon interference. We use the general expression for the event probabilities~\cite{Tichy15}:
\begin{equation}
    P_M=\sum_{\rho,\sigma\in S_N}\prod_{j=1}^N M_{\sigma_j,j}M^*_{\rho_j,j}x_{\sigma_j,\rho_j}
    \label{partialeq}
\end{equation}
where $N$ is the total number of photons, $M$ is an $N$x$N$ submatrix of the transfer matrix corresponding to a particular choice of $N$ input and output modes, and $x_{j,k}$ is the overlap parameter between the $j$th and $k$th photons. $\rho$ and $\sigma$ refer to permutations of the indices 1 to $N$, and are summed over the symmetric group $S_N$. One obvious case which contains 2 photon but not 3 photon interference is that where one of the photons is completely distinguishable but the others interfere perfectly. Conversely, it may be more accurate to consider some uniform overlap $x$ between any pair of the photons. Following \cite{Renema18}, in this case the probability can be expressed as a series in $x$:
\begin{equation}
    P_M=P_M^{(0)}+x^2P_M^{(2)}+x^3P_M^{(3)}
\end{equation}
where $x^kP_M^{(k)}$ can loosely be thought of as the contribution from $k$ photon interference, with $P_M^{(0)}$ the probability for completely distinguishable photons. For $x<1$ the higher-order contributions are increasingly attenuated, and truncating the series at some $k$ may provide an accurate approximation while being easier to compute classically than the full model; in this case, neglecting $P_M^{(3)}$ while keeping $P_M^{(2)}$.

Fig.~\ref{fig3}(b) shows the distances of the 3 photon data to the various theoretical models. To estimate uncertainties we simulate Poissonian count statistics many times, calculating $D$ each time then taking its mean and standard deviation. This generally increases $D$ compared to its raw value. (i) uses the ideal distribution with perfect overlaps. (ii) includes a uniform overlap parameter $x$ while retaining all orders of interference. By optimising over $x$, a slight decrease in $D$ is seen. (iii) is based on the truncated model, effectively keeping 2 photon interference but neglecting 3 photon interference. (iv) to (vi) are the three cases where two of the photons interfere perfectly but the other one is completely distinguishable from them. (vii) is the case where all photons are completely distinguishable from each other. The models (iii)-(vii) do not contain 3 photon interference, and all have significantly larger $D$ than (i) or (ii), so we can tentatively conclude that our results do contain significant contributions from 3 photon interference.

The degree of distinguishability between photons is largely a measure of the quality of the states generated by the SPDC sources - however the observation of multi-photon interference also confirms that the interferometer functions correctly and does not introduce distinguishability. For example the internal path lengths could differ, or the DC reflectivities could vary with wavelength over the bandwidth of the photons. It has also been seen that multi-mode interferometer devices, which are often used as beamsplitters in silicon photonics, can introduce temporal distinguishability between broadband photons~\cite{Peruzzo11}. It can be shown~\cite{supp} that such distinguishability induced by the circuit can be incorporated into the same expression for the event probabilities as in Eq.~\ref{partialeq}.

Four photons were then injected into the same interferometer, by connecting the four channels from the sources to the four input modes. There are multiple ways to generate a four photon event in this configuration: each source can generate a pair of photons, or either of the sources can generate two pairs, all with equal probability. We assume that the phases between these different contributions vary randomly over the time of the experiment, due to fluctuations in the bulk optics around the sources and the fibers connecting the source setup to the interferometer, such that we can treat the input state as an incoherent mixture of the three contributions. The experimental distribution over the 715 output combinations, shown in Fig.~\ref{fig4}(a) after 5 hours of integration at a rate of 0.8Hz, has a fidelity $F=0.97$ and a raw distance $D=0.13$ to the ideal case.

Noting that in the three photon case, the truncated series expansion model came closest to describing the results without including three photon interference (i.e. model iii came closer to describing that data than models iv-vii), we now compare the four photon results to this model with different levels of truncation. Fig.~\ref{fig4}(b) shows the distances to the different models, with (i) the ideal case of indistinguishable photons and (ii) the model with partial distinguishability but including all orders of interference. (iii) is the truncated series model neglecting four photon interference but keeping lower orders; (iv) is the same but also neglecting three photon interference. (v) is the case of completely distinguishable photons. With the exception of (i) and (v), the overlap parameter $x$ has been optimised separately for each case to minimise the distance. It can be seen that this process improves the distance to the experimental results for (ii) compared to (i). Then for each successive level of approximation the distance is increased, suggesting that all orders of interference up to 4 photons are required to describe the experimental data. However, the increase with each level of approximation is less dramatic than in the 3 photon case. The full model with partial distinguishability is slightly further from the experimental results than before, which is probably explained by various errors having a larger effect, such as slight uncertainties in the characterization of the transfer matrix, non-uniform overlap parameters, imperfect spectral purity, and statistical errors. More surprisingly, the 2nd order truncated model which contains $P_M^{(0)}$ and $P_M^{(2)}$ comes closer to the experiment for 4 photons than for 3 - it may be that using an incoherent mixture of 4 photon input states washes out higher-order interference effects to some extent, so that the truncated models become better approximations.

Finally, we test 5 photon interference, now using the 15 mode interferometer. Here, source A generates 2 photon pairs, such that there are 2 photons at both input 1 and input 2; meanwhile source B generates a heralded single photon which is injected into input 4. There are 3003 possible output combinations, and combined with the reduced event rate of a few per hour this makes it impractical to accumulate an accurate picture of the output distribution. The low event rate is not solely due to the reduced probability of generating and detecting higher numbers of photons - for the case of ideal interference, only 15\% of events are expected to be collision-free with each photon arriving at a different output port. Since we use threshold detectors, the $\sim 85\%$ of events with 2 or more photons sharing an output are not registered as coincidences. This could be remedied by using photon-number resolving detectors, but in general it is the collision-free cases that are hardest to simulate classically~\cite{aaronson11}. This demonstrates the need to increase the number of interfering modes in order to have meaningful boson sampling with >5 photons. Instead of calculating the distance or fidelity of the output to a model distribution, we consider the likelihood ratio, defined here as the direct ratio between the probability that the set of samples $\vec{S}$ came from a model $A$ or a model $B$:
\begin{equation}
    L=\frac{P(\vec{S}|A)}{P(\vec{S}|B)}.
\end{equation}
Below we set model $B$ to be the ideal case of indistinguishable photons, and set $A$ to be various alternatives that do not contain 5 photon interference, such that a value below 1 suggests the photons are interfering correctly.

\begin{figure}[t!]
	\centering
	\includegraphics[width=0.7\columnwidth]{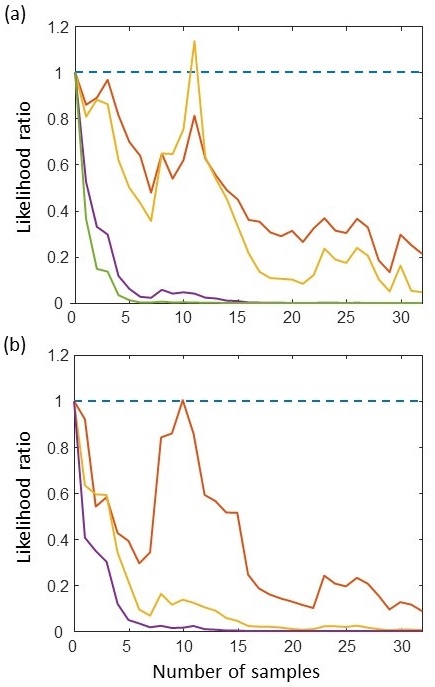}
        \vspace{-10pt}
	\caption{5 photon interference likelihood ratios for various models. Values <1 indicate the measured samples were more likely to have come from the ideal case. (a) green line: fully distinguishable photons; purple: series expansion truncated at 2 photon interference; yellow: truncated at 3 photon interference; red line: truncated at 4 photon interference. (b) red line: one of the photons in input 1 is distinguishable, the others interfere perfectly; yellow line: one of the photons in input 2 is distinguishable; purple line: the photon in input 4 is distinguishable.}\label{fig5}
 \vspace{-5pt}
\end{figure}

In Fig.~\ref{fig5}(a) we plot $L$ as a function of the number of samples for different truncations of the series expansion. It can be seen that the line corresponding to completely distinguishable photons decays very rapidly, and that after a few samples $L$ is close to zero. Similarly the model containing 2 photon interference contributions has decayed after around 15 samples. The models containing 3 and 4 photon interference perform better, but over 30 samples can clearly be seen to be decaying. In Fig.~\ref{fig5}(b) we consider the models where just one of the photons is distinguishable from the others - the cases with a distinguishable photon at input 2 or 4 decay rapidly, while the case with a distinguishable photon at input 1 does better but has still decayed significantly over 30 samples.

In all cases, the results are more likely to have come from genuine multi-photon interference than from any of the alternatives. This clearly does not rule out all possible models which do not contain 5 photon interference - it may well be possible to find a better fit to the experimental data by varying more free parameters, such as more general choices of the overlap parameters between photons, or by changing the unitary matrix. Nonetheless we have ruled out a selection of physically motivated models which use the transfer matrix obtained from seemingly reliable single photon and two photon measurements, and which consider the representative cases of either a uniform level of distinguishability between photons or just one photon being entirely distinguishable from the others. In particular, when we consider the series expansion of the event probability with uniform distinguishability, we see that including successive terms in the series, corresponding to higher orders of interference, leads to an increasingly good approximation to the data. While it does not have to be the case at the level of individual event probabilities that including more terms monotonically improves the approximation, it is reassuring to see that this is the case in the aggregate, with each term leading to an appreciable improvement in the distance or likelihood ratio to the data.

In conclusion, we have demonstrated interference between 3, 4, and 5 photons in integrated silicon photonic circuits. The photons were generated off-chip using SPDC in ppKTP waveguides, which provide a bright, efficient, and reproducible source of pure and indistinguishable photon pairs. The usual difficulty in interfering photons from SPDC sources in silicon circuits is the low transmission through the chip - this was obviated here by the use of high-efficiency grating couplers to get the photons on- and off- chip, and by a compact and low-loss interferometer design. This has enabled 5 photon interference in a 15 mode Haar-random interferometer. However, the rate of 5 photon events remains low in comparison to boson sampling with bulk-optic interferometers and the most efficient sources available~\cite{Wang17, Zhong18}, which emphasises the need to further reduce loss, and to increase the number of sources, which would improve the generation rate for higher photon numbers in a scattershot or Gaussian boson sampling scenario.

\section{Funding}

The Networked Quantum Information  Technologies  Hub (NQIT)  as  part  of  the UK National Quantum Technologies Programme Grant (EP/N509711/1); the Key R$\&$D Program of Guangdong Province (2018B030325002); the Local Innovative and Research Teams Project of Guangdong Pearl River Talents Program (2017BT01X121); the Guangzhou Science and Technology Plan Project(201707010096). G.S.T. acknowledges financial support from the Natural Sciences and Engineering Research Council of Canada and the Oxford Basil Reeve Graduate Scholarship.

\section{Acknowledgment}

The authors thank Jelmer Renema for useful discussions and Andreas Eckstein for his help with the source setup.




\newpage
\onecolumngrid
\section*{\fontsize{14}{21}\selectfont Supplementary Information}
\section*{Fabrication flow}
\begin{figure*}[h]
	\centering
	\includegraphics[width=0.7\columnwidth]{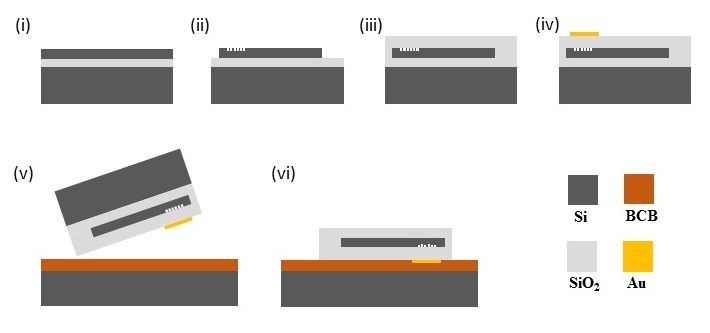}
        \vspace{-10pt}
	\caption{The fabrication flow.}\label{figS1}
 \vspace{-5pt}
\end{figure*}

The fabrication flow is shown in Fig.~\ref{figS1}. (i) The device was fabricated on a 220 nm silicon-on-insulator (SOI) sample. (ii) First, standard SOI processing, including e-beam lithography and inductively coupled plasma etching processes, was used to fabricate the silicon waveguides and grating couplers. (iii) Then, a thickness of 3 $\upmu$m SiO$_2$ layer was coated on the top of the fabricated device and plannarized by chemical mechanical polishing, resulting in a flat surface. The thickness of the SiO$_2$ needs to be controlled carefully to optimize the grating coupler, because it will define the distance from the grating coupler to the bottom mirror, and the field reflected from the bottom mirror should interfere constructively with that directly diffracted by the grating coupler~\cite{Luo18}. (iv) Afterward, 5 nm of Ti and 100 nm of Au were patterned right above the grating coupler to form the bottom mirror. (v) The sample was then flip-bonded to a silicon carrier wafer through a chip to wafer bonding process using benzocyclobuten (BCB). (vi) Finally, the substrate of the chip was removed by mechanical grinding and dry etching.

\section*{Distinguishability introduced by a linear optical circuit}

The effects of partial distinguishability between photons in multiphoton interference have been investigated thoroughly, both in theory and experiment. Normally it is assumed that the partial distinguishability exists in the photons at the start of the interferometer, and is most likely a property of the single photon source. Less attention has been given to imperfections in the interferometer which could introduce distinguishability; for instance mismatched internal path lengths or group-velocity dispersion could create temporal distinguishability, stray birefringence could rotate the photons' polarization, and the beamsplitter ratios could be frequency dependent, creating correlation between a photon's path through the interferometer and its spectral state. Below, we show that these imperfections can conveniently be incorporated into the tensor-permanent form used for partially distinguishable photons~\cite{Tichy15}, and hence that this model is a very general way to describe non-ideal multi-photon interference.

We consider the event probability for photons in input ports 1 to $N$ to be found at the output ports 1 to $N$ given a transfer matrix $M_{j,k}^{\omega,\omega'}$. This transformation acts not just on the port degree of freedom, but on an internal degree of freedom $\omega$. The input state contains single photons in modes 1 to $N$, with internal states $f_j(\omega)$.
\begin{equation}
    \ket{\psi_{\text{in}}}=\prod_{j=1}^N \left(\int d\omega ~f_j(\omega)a_{j,\omega}^\dagger\right) \ket{0}.
\end{equation}
After the circuit, the component of the state with photons at output ports 1 to $N$ is given by
\begin{equation}
    \ket{\psi_\text{out}}=\sum_{\sigma\in S_N}\prod_{j=1}^N\left(\iint d\omega d\omega' ~M_{\sigma_j,j}^{\omega,\omega'}f_{\sigma_j}(\omega)a^\dagger_{j,\omega'}\right)\ket{0},
\end{equation}
where we sum over all permutations of the photons between input and output, and apply the permutation to the input port indices, such that $f_j(\omega)$ becomes $f_{\sigma_j}(\omega)$. Then take the projection onto an output state with a photon at each port with internal degree of freedom $\Omega_j$:
\begin{equation}
    \bra{\phi}=\bra{0}\prod_{j=1}^N a_{j,\Omega_j}
\end{equation}
\begin{equation}
    \braket{\phi|\psi_\text{out}}=\sum_{\sigma\in S_N}\prod_{j=1}^N\iint d\omega d\omega' ~M_{\sigma_j,j}^{\omega,\omega'}f_{\sigma_j}(\omega)\delta(\Omega_j-\omega')=\sum_{\sigma\in S_N}\prod_j\int d\omega ~M_{\sigma_j,j}^{\omega,\Omega_j}f_{\sigma_j}(\omega).
\end{equation}
The overall event probability is the absolute square of the overlap, integrated over all the $\Omega$ to trace out the internal degrees of freedom:
\begin{equation}
    P=\int d\vec{\Omega} \sum_{\sigma,\rho\in S_N}\prod_{j=1}^N  \left(\int d\omega~M_{\sigma_j,j}^{\omega,\Omega_j}~f_{\sigma_j}(\omega)\right) \left(\int d\omega~ M_{\rho_j,j}^{\omega,\Omega_j}~f_{\rho_j}(\omega)\right)^*.
\end{equation}
Since each term inside the summation and product only depends on one of the $\Omega_j$s, this integral can also be moved inside:
\begin{equation}
    P=\sum_{\sigma,\rho\in S_N}\prod_{j=1}^N\int d\Omega   \left(\int d\omega~M_{\sigma_j,j}^{\omega,\Omega}~f_{\sigma_j}(\omega)\right) \left(\int d\omega~ M_{\rho_j,j}^{\omega,\Omega}~f_{\rho_j}(\omega)\right)^*.
\end{equation}
This can be expressed as a tensor permanent:
\begin{equation}
    P=\sum_{\sigma,\rho\in S_N}\prod_{j=1}^N~W_{\sigma_j,\rho_j, j}~~~~~~~~~~~~~~~W_{j,k,l}=\int d\Omega \left(\int d \omega~M_{j,l}^{\omega,\Omega}f_j(\omega)\right) \left(\int d \omega~M_{k,l}^{\omega,\Omega}f_{k}(\omega)\right)^*.
\end{equation}
$W$ is the inner product of two unnormalised internal states which incorporate the effect of the linear optical transformation:
\begin{equation}
    W_{j,k,l}=\int d\Omega~ V_{j,l}^\Omega~V_{k,l}^{\Omega *}~~~~~~~~~~~~~~~V_{j,k}^\Omega=\int d \omega~M_{j,k}^{\omega,\Omega}f_j(\omega).
\end{equation}
$V_{j,k}^\Omega$ is the amplitude associated with having a single photon at output $k$ and frequency $\Omega$ given one input photon at port $j$ with spectrum $f_j(\omega)$. There is some redundancy in the description of an input internal state $f_j(\omega)$ and a circuit which can operate on the internal state - we could as easily start from any generic set of internal states, then incorporate transformations acting on the internal state at the start of the circuit to map it to the true state. The state after the circuit could more simply have been written in terms of $V$:
\begin{equation}
    \ket{\psi_\text{out}}=\sum_{\sigma\in S_N}\prod_{j=1}^N\left(\int d\omega ~V_{\sigma_j,j}^\omega~a^\dagger_{j,\omega}\right)\ket{0},
\end{equation}

$W_{j,j,k}$ is the probability of a single photon transitioning from port $j$ to $k$ with the internal state traced out, which is real and non-negative:
\begin{equation}
    W_{j,j,k}=\int d\Omega~ |V_{j,k}^\Omega|^2.
\end{equation}

The other elements of $W_{j,k,l}$ are complex numbers which contain the relative phase and a measure of the coherence between the two transitions $j\rightarrow l$ and $k\rightarrow l$. We have that $W_{j,k,l}=W_{k,j,l}^*$. Unlike a general description of a quantum channel, $W$ does not contain information about the coherence between transitions with different output ports - this is because we assume the state is measured directly after the transformation, so this coherence is not relevant. By the Cauchy-Schwarz inequality,
\begin{equation}
    |W_{j,k,l}|\leq\sqrt{W_{j,j,l}W_{k,k,l}}.
\end{equation}
This inequality is saturated, i.e. $|W_{j,k,l}|=\sqrt{W_{j,j,l}W_{k,k,l}}$, when the photons are identical and the unitary has no dependence on the internal state, which is a return to the ideal case of perfect interference. This shows that the effect of the circuit on the internal states and the dependence of the transfer matrix on the internal states can only act to reduce these coherences (and potentially change their phase) relative to the incoherent transition probabilities, in the same manner as partial distinguishability between photons does.

\end{document}